# Dispersive coherent Brillouin scattering spectroscopy

Ayumu Ishijima[1,2,*], Shinga Okabe[3], Ichiro Sakuma[1,3,4], and Keiichi Nakagawa[1,3]

[1]*Department of Precision Engineering, The University of Tokyo, Tokyo 113-8656, Japan.*
[2]*PRESTO, Japan Science and Technology Agency, Saitama 332-0012, Japan*
[3]*Department of Bioengineering, The University of Tokyo, Tokyo 113-8656, Japan.*
[4]*Medical Device Development and Regulation Research Center, The University of Tokyo, Tokyo 113-8656, Japan*
[*]*Corresponding author: ishijima@bmpe.t.u-tokyo.ac.jp*

Ayumu Ishijima (ishijima@bmpe.t.u-tokyo.ac.jp, +81-3-5841-2931)
Shinga Okabe (okabe@bmpe.t.u-tokyo.ac.jp)
Ichiro Sakuma (sakuma@bmpe.t.u-tokyo.ac.jp)
Keiichi Nakagawa (kei@bmpe.t.u-tokyo.ac.jp)

# Abstract

Frequency- and time-domain Brillouin scattering spectroscopy are powerful tools to read out the mechanical properties of complex systems in material and life sciences. Indeed, coherent acoustic phonons in the time-domain method offer superior depth resolution and a stronger signal than incoherent acoustic phonons in the frequency-domain method. However, it does not allow multichannel detection and, therefore, falls short in signal acquisition speed. Here, we present Brillouin scattering spectroscopy that spans the time and frequency domains to allow the multichannel detection of Brillouin scattering light from coherent acoustic phonons. Our technique maps the time-evolve Brillouin oscillations at the instantaneous frequency of a chromatic-dispersed laser pulse. The spectroscopic heterodyning of Brillouin oscillations in the frequency domain enhances the signal acquisition speed by at least 100-fold over the time-domain method. As a proof of concept, we imaged heterogeneous thin films and biological cells over a wide bandwidth with nanometer depth resolution. We, therefore, foresee that our approach catalyzes future phonon spectroscopy toward real-time mechanical imaging.

# 1. Introduction

The characterization of the micro/nanoscale material's mechanical properties is critical for an understanding of their functionality[1,2]. Researchers have been looking for a means to measure on a tiny scale for several decades. However, very few tools are available for obtaining the mechanical properties of miniaturized samples including biological cells. Common mechanical tests (e.g., dynamic mechanical analysis and rheometer) used for bulk materials require contact mechanical forces on samples, and it is difficult to scale down their size to obtain small-scale spatial information. Although atomic force microscopy[3,4] enables two-dimensional (2D) measurements with a resolution up to nanometers, it is limited to surface measurements and requires the contact of the indenters with the specimens. Ultrasound elastography[5,6] provides access to mechanical properties in a non-invasive fashion but lack micrometer-level resolution. Thus, despite the advancements achieved in mechanical testing techniques, the realization of in-depth, noncontact microscopic measurements remains an open challenge.

An acousto-optical technique can meet this demand: Brillouin scattering spectroscopy interrogates material's mechanical properties from an inelastic light scattering by a phonon, referred to as Brillouin light scattering[7]. It non-invasively reads out the viscoelastic properties of small-scale materials with high spatial resolution[8–10]. Brillouin scattering spectroscopy has various applications, including phonon physics studies[11–13] and characterization of solids[14–16], liquids[17,18], and biological samples[19–24]. In general, Brillouin scattering spectroscopy can be classified into two forms: (1) as frequency-domain Brillouin scattering spectroscopy, in which gigahertz-frequency-shifted Brillouin scattering light from incoherent acoustic phonons is recorded using a specific imaging spectrometer (e.g., virtually imaged phased array spectrometer[25]), and (2) as time-domain Brillouin scattering spectroscopy, in which the carrier frequency of Brillouin scattering light from coherent acoustic phonons, known as Brillouin oscillation, is recorded by coherent Brillouin scattering techniques (referred to as picosecond ultrasonics[18,26,27]) typically based on the step-scan pump-probe method with a single photodetector and an optical delay line. Although coherent acoustic phonons in the time-domain method offer superior depth resolution and a stronger signal when compared with incoherent acoustic phonons in the frequency-domain method, it does not allow multichannel detection and, therefore, falls short in the signal acquisition speed. This issue is an open challenge for the Brillouin scattering spectroscopy to extend it from point sampling to nano-imaging modalities.

In this article, we propose and demonstrate Brillouin scattering spectroscopy that straddles the time and frequency domains, which we refer to as dispersive coherent Brillouin scattering spectroscopy. Our technique harnesses coherent acoustic phonons and a chirped laser pulse to record time-evolving Brillouin oscillations in the frequency domain. Coherent acoustic phonons launched from a metallic thin film provide nanometer-scale depth resolution and intense Brillouin scattering light, as in time-domain Brillouin scattering spectroscopy. The chirped laser pulse with linear chromatic dispersion gives multichannel detection of Brillouin scattering light, as in frequency-domain Brillouin scattering spectroscopy. Hence, our technique inherently possesses a higher depth resolution and better signal-to-noise ratio than conventional frequency-domain Brillouin scattering spectroscopy and at least two orders higher signal acquisition speed than conventional time-domain Brillouin scattering spectroscopy. Our dispersive coherent Brillouin scattering spectroscopy is as versatile as conventional Brillouin scattering spectroscopy and holds great promise as a practical microscopic mechanical imaging technique.

## 2. Results

### 2.1. Working principle of dispersive coherent Brillouin scattering spectroscopy

In our technique of dispersive coherent Brillouin scattering spectroscopy, we use a single ultrafast short-pulse laser to pump and probe coherent acoustic phonons (**Figure 1**a). Coherent acoustic phonons are launched into the sample by the absorption of an optical pump pulse and the subsequent rapid thermal expansion of a metallic thin film, which acts as a transducer (Figure 1b). A femtosecond laser pulse is temporally stretched by the pulse stretcher (left inset in Figure 1a) and made incident on the sample as successive probes diffracted by propagating coherent acoustic phonons at different timing and positions (Figure 1b). In other words, the frequency of the probe photons encodes the time delay which is typically set using a mechanical delay stage that must be scanned. Heterodyning the diffracted chirped laser pulse from coherent acoustic phonons and the reflected chirped laser pulse from the surface of the transducer leads to the gigahertz-beating of the light intensity in the time domain. A spectrometer (right inset in Figure 1a) disperses the chirped laser pulse in the frequency domain to create a Brillouin interferogram (Figure 1c). The frequency $f_B$ of these oscillations is related to acoustic velocity $v$, refractive index $n$, and probe wavelength $\lambda$ through the relation $f_B = 2nv/\lambda$ for normal probe beam incidence. The Brillouin interferograms recorded by the image sensor are processed on a computer to construct a time trace of transient reflectivity with the time axis calibrated based on the optical settings (Figure S1, Supporting Information). A high sampling frequency (4.2 THz) and an ample time window (282 ps) can be achieved with the large number of data points (1024 points). The temporal resolution of the system is determined by the sub-pulse duration[28] given by $(\tau_0 \tau_c)^{1/2}$, where $\tau_0$ is the Fourier transform-limited pulse duration and $\tau_c$ is the chirped pulse duration. The maximum time window is traded off with the temporal resolution, where increasing the chirp elongates the time window but degrades the temporal resolution (Figure S2, Supporting Information). We used a chirped pulse of a full-width at half-maximum (FWHM) temporal duration of 160 ps and 400 nm-centered wavelengths with an FWHM bandwidth of 7.8 nm corresponding to a 31 fs transform-limited pulse duration. Based on these parameters, our current system has a temporal measurement resolution of 2 ps. Furthermore, the probe bandwidth of 7.8 nm used in this study is narrow enough for probing photons to "see" the same phonon wavepacket in the linear range of dispersion[29].

## 2.2. Basic performance of dispersive coherent scattering Brillouin spectroscopy

To gauge the performance of dispersive coherent scattering Brillouin spectroscopy, we measured the Brillouin frequency of a silicon dioxide ($SiO_2$) thin film attached to a 100 nm chromium film (**Figure 2**a). We used $SiO_2$ thin film as the sample for our proof-of-principle demonstration because the acoustic velocity ($v \sim 5.9$ km/s), refractive index ($n \sim 1.47$), and Brillouin frequency ($f_B \sim 44$ GHz) at $\lambda \sim 400$ nm were reported in earlier studies[30,31]. The time trace of the Brillouin oscillations (Figure 2b and Figure S3, Supporting Information) and the Brillouin spectrum (Figure 2c) obtained using our technique (averaged over 4000 shots, which corresponds to a data acquisition time of 40 s in our experimental setup) are in good agreement with those measured using time-domain Brillouin scattering technique (step-scan method). The thickness of the $SiO_2$ thin film calculated from a bump on the time trace around 180 ps originating from the $SiO_2$–air interface is 1 μm, which matches the actual thickness of the $SiO_2$ film. Further, we showed that our technique can measure the Brillouin frequency of liquid and gel samples (Figure S4, Supporting Information). Meanwhile, there were slight differences in the Brillouin oscillation amplitude and a non-oscillatory signal component owing to thermally induced changes in the reflectivity of chromium film originating from indispensable wavelength dependence on piezo-optical couplings[32]. It is also worth mentioning that there is more interest in the Brillouin frequency than the transient changes in the refractive index of the transducer, which are routinely removed[33].

We show the nanometre-resolution depth profiling of a heterogeneous thin film consisting of silicon nitride ($Si_3N_4$) and $SiO_2$ (Figure 2d); $Si_3N_4$ is widely used in optoelectrical devices, and its Young's modulus is approximately three times higher than that of $SiO_2$[31]. From the time trace presented in Figure 2e, one can recognize the change in the phase-matching condition of the probe light and coherent acoustic phonons when the coherent acoustic phonons leave $Si_3N_4$ and enter $SiO_2$. High-frequency oscillations ($f_B$ = 90 GHz) from the $Si_3N_4$ layer appear first, and low-frequency oscillations ($f_B$ = 44 GHz) start with a delay of 80 ps, which corresponds to the crossing of coherent acoustic phonons at the interface of the $SiO_2$ layer (Figure 2f). $Si_3N_4$ has a higher Brillouin frequency than $SiO_2$, as expected, because the product of the material's refractive index and acoustic velocity differs by a factor of 2.

We also performed the depth profiling of a 1μm $SiO_2$ step created on a 300 nm titanium film (Figure 2g). We obtained spatial profile by translating the sample over 3 mm with a 100 μm pitch. From the transient reflectivity map shown in Figure 2h, we can observe Brillouin oscillations from the $SiO_2$ layer up to a lateral position of 2.2 mm. The thickness of the $SiO_2$ thin film (Figure S5, Supporting Information) corresponds well to the bumps on the time trace around 150 ps originating from the $SiO_2$–air interface. In this context, the timing where coherent acoustic phonons reached the interface between titanium and $SiO_2$ ($T_1$) and $SiO_2$ and its surface ($T_2$) were defined as jumps in the transient reflectivity waveforms. The spatial profile of the film thickness was calculated using the time integration of the acoustic velocity $v$ between $T_1$ and $T_2$ at each measurement position. The lateral boundary of the film becomes apparent near 1.2 mm, and the thickness of the film decreases as it approaches the bare titanium region (Figure 2i). Furthermore, we demonstrate a cross-sectional Brillouin frequency mapping capability. We used glycerol, a well-known and well-characterized prototypical glass-forming liquid[34], squeezed between two flat glass substrates; one of them was a 300 nm titanium film with a 1 μm $SiO_2$ step. We can observe the interface between $SiO_2$ and glycerol from the contrast produced by the Brillouin frequency (Figure S6, Supporting Information), whereas their close refractive index values ($n \sim 1.47$ and 1.48 for $SiO_2$ and glycerol, respectively) make it difficult to see the interface with the naked eye.

## 2.3. Three-dimensional Brillouin imaging by dispersive coherent Brillouin scattering spectroscopy

To demonstrate the 3D Brillouin imaging capability, we imaged phosphate-buffered saline squeezed between two flat glass substrates; one of them was a 300 nm titanium film with a 1 μm $SiO_2$ step (**Figure 3**a). We tuned the time window to 844 ps from 282 ps for improving the low Brillouin frequency resolution and enlarging the depth direction field of view by simply adjusting the distance between the grating pairs (Figure S7, Supporting Information). The temporal measurement resolution was 3.8 ps in this configuration (Figure S2, Supporting Information). 3D transient reflectivity maps were obtained by raster scanning the sample in two dimensions (fixed $z$ axis) with a 5 μm step over 100 μm, averaging over 2000 pairs of measurements without and with the pump (0.9 mJ/cm$^2$) for each step. We can observe the interface between $SiO_2$ and phosphate-buffered saline from the contrast produced by the Brillouin frequency (Figure 3b–3d). The spatial variation in the Brillouin frequency at the calculated depth positions is mapped, as depicted in Figure 3e (Video S1, Supporting Information). The area occupied by $SiO_2$ in the cross-sectional image decreases with increasing height because of the slope set in the sample.

The microscopic imaging capabilities were further illustrated using biological cell samples (**Figure 4**a and Figure S8, Supporting Information). We used HeLa cells fixed on a 300 nm titanium film, and the sample was scanned in two dimensions with a 3 µm step over 60 µm to acquire 3D transient reflectivity maps. We averaged over 2000 pairs of measurements without and with the pump (0.04 mJ/cm$^2$) for each step, resulting in an acquisition time of 40 s/step. The Brillouin frequency image clearly shows the difference in the Brillouin frequency of the internal structure, particularly the nucleus (yellow region) and cytoplasm of the cell (Figure 4b–4d). The spatial variation in the Brillouin frequency at the calculated depth positions is mapped in Figure 4e (Video S2, Supporting Information). The high-frequency regions around the cells in the shallow area originate from the surface displacement of the Ti film and not the frequencies originating from the Brillouin oscillations. Therefore, these regions were masked for improved visualization. Furthermore, we acquired the spectral images by plotting the spectral magnitude of a distinguishable frequency signature of the sample to clearly visualize the cellular structure in three dimensions (Figure 4f and Video S3, Supporting Information).

## 3. Discussion and Conclusion

In this study, the feasibility of dispersive coherent Brillouin scattering spectroscopy is demonstrated. The technique combines two key mechanisms—localized spectral encoding of the Brillouin oscillation and decoding by a spectrometer—to perform the multichannel detection of Brillouin oscillations. Consequently, our technique can acquire Brillouin oscillations at least 100-times faster than the step-scan method used in conventional time-domain Brillouin scattering spectroscopy. The abovementioned comparisons were performed under the following conditions: identical laser repetition rate, measurement time window, and temporal resolution and an optical delay line with an infinitely short positioning time. In this case, the proposed method reduces the measurement time by the number of scanning points of the optical delay line. Therefore, we achieved a 126-fold (222-fold for imaging experiments) improvement in the signal acquisition speed, given that the temporal resolution was 2.2 ps (3.8 ps) and the measurement time window was 282 ps (844 ps).

As of the current proof-of-principle experimental setup, the acquisition speed does not surpass asynchronous optical sampling (ASOPS) systems[35–38] that improved the acquisition rate to a few seconds per point. ASOPS systems rapidly scan delay-time by employing two synchronized mode-locked pulsed lasers with a slight repetition rate offset. However, adopting two ultrafast pulsed laser sources increase the complexity requiring phase-controlling devices and stabilization electronics. In addition, the need for precisely controlling the repetition rates of two lasers trades mechanical complexity for electronic complexity. Moreover, ASOPS approaches are impractical for experiments that use high-power laser sources with large shot-to-shot fluctuations or low repetition rates.

Although the frame interval of the image sensor used as a multichannel detector limits the overall measurement time of the current proof-of-principle experiment, this limitation can be addressed through further system improvements from numerous directions, such as using a high-frame-rate camera with a higher repetition rate laser source. Changing the operating wavelength to the near-infrared region to employ optical fiber-based dispersive Fourier-transform spectroscopic techniques[39,40] would further improve the acquisition rate of our method. Moreover, the measurement time window can be easily tuned to nanosecond time scales by the dispersion in the optical fiber. The optical fiber-based system can profoundly benefit from translating our technique toward a study on moving biological specimens and flow cytometric applications. Thus, its potential applications are versatile and introduce a new paradigm in Brillouin scattering spectroscopy to open a door for the prospect of unexploited fields in material and life sciences.

## 4. Experimental Section

### 4.1. Brillouin scattering spectroscopy system

The ultrashort laser source we used was a Ti:sapphire regenerative amplifier (Astrella-USP-1K, Coherent, US) producing 803 nm-centered wavelengths with a bandwidth of 70 nm, and a pulse duration of 35 fs amplified pulses in a maximum repetition rate of 1 kHz. Because the sensor readout used in the spectrometer led to slow data acquisition, the pulse repetition rate was down counted from 1 kHz to 100 Hz. The laser output was frequency-doubled (406 nm-centered wavelengths with a bandwidth of 7.8 nm) by a $\beta$-barium borate crystal (BBO-1001H, Eksma Optics, LT) and separated from an 800 nm laser pulse by a harmonic separator (USB21, Thorlabs, US) to use as a probe pulse.

An 800 nm fs laser pulse, modulated at 50 Hz by an optical chopper (MC2000B, Thorlabs), served as a pump pulse to excite coherent acoustic phonons. A probe pulse was time-stretched to by a pair of 36000 lines/mm gratings (PC3600, Spectrogon, SE). Subsequently, the probe pulse was split into a sample and a reference arm by a beam splitter (BS013, Thorlabs). After passing via a half-wave plate (WPH10M-405, Thorlabs), polarizing beam splitter (PBSW-405, Thorlabs), and a quarter-wave plate (SAQWP05M-700, Thorlabs) in the sample arm, a lens ($f$ = 40 mm) focused a probe pulse on the sample. For microscopy experiments, we changed the lens to an objective lens with a numerical aperture of 0.42 (M-Plan Apo 20×, Mitsutoyo, JP). The probe pulse reflected from the sample, and a fraction of probe pulse that bypassed the sample was directed into a homebuilt Czerny–Turner spectrometer (spectral resolution of 0.037 nm, pixel resolution of 0.0169 nm/pixel, spectral bandwidth of 17.3 nm) coupled to a 16-bit sCMOS camera (Orca Flash4.0 V3, Hamamatsu Photonics, JP) by a cylindrical lens ($f$ = 200 mm) for realizing balanced detection. Our spectrometer consisted of two spherical mirrors (CM508-500-F01, Thorlabs) and a 2400 lines/mm grating (GH50-24V, Thorlabs). Using a digital delay generator (DG645, Stanford Research Systems, US), the camera and the optical chopper (MC2000B, Thorlabs) were synchronized to the laser's repetition rate. The samples were mounted on motorized stages connected to a stage controller (SHOT304GS, Sigmakoki, JP).

### 4.2. Transient reflectivity calculation

Transient reflectivity waveforms were constructed from the captured images, yielding a sensitivity of approximately $10^{-5}$ with $10^3$ laser shots (Figure S9, Supporting Information) by following procedures (Figure S10a, Supporting Information). First, we excluded pixels at or near the boundary of signal and reference regions. Second, each region corresponding to a specific delay time was vertically averaged over roughly 500 pixels. Finally, we normalized the value based on the signal ($I_{sig}$) and reference ($I_{ref}$) spectra recorded with ($w$) and without ($wo$) the pump pulse:

$$\frac{\Delta R}{R} = \frac{I_{sig}^{w}}{I_{ref}^{w}} \bigg/ \frac{I_{sig}^{wo}}{I_{ref}^{wo}} - 1 \qquad (1)$$

### 4.3. Time-domain Brillouin scattering spectroscopy with step-scan method

We used an optical setup similar to that described above without the pulse stretcher and spectrometer. Two photodiodes (SM1PD1A, Thorlabs) with current preamplifiers (PDA200C, Thorlabs) were used instead of the spectrometer. The peak voltage output from the current preamplifier was recorded using a data acquisition card (USB-6216, National Instruments, US) with an external sampling clock produced by a delay generator (DG535, Stanford Research

Systems, US) to measure the pulse energy of each probe pulse[41]. The delay time between pump and probe pulses was changed by stepping the optical delay line set in the pump arm.

### 4.4. Brillouin imaging

Transient reflectivity waveforms at each measurement position of the sample were acquired by translating the sample mounted on a three-axis motorized stage connected to a stage controller. We produced a Brillouin frequency map by processing each waveform with continuous wavelet transformation using the generalized Morse wavelet to obtain the peaks of Brillouin frequency $f_B$ at different timing (Figure S10b and S10c, Supporting Information). The frequency domain representation of the Morse wavelet[42] is

$$\Psi_{P,\gamma}(\omega) = U(\omega) a_{P,\gamma} \omega^{\frac{P^2}{\gamma}} \exp(-\omega^\gamma) \qquad (2)$$

where $U(\omega)$ is the unit step function, $a_{P,\gamma}$ is a normalizing factor, $P^2$ is the time-bandwidth product, and $\gamma$ is the asymmetry parameter. We used the symmetry parameter $\gamma$ of 3, and the time-bandwidth product $P^2$ of 5 and 60 for $SiO_2$ and the cell sample, respectively. The depth profile of the sample was constructed by calculating the acoustic velocity $v$ from the Brillouin frequency map through the relation of $v = f_B \lambda / 2n$. The refractive index $n$ for each sub-pulse wavelength $\lambda$ was calculated from the dispersion formula[43,44] for the $SiO_2$ sample. The refractive index $n$ for phosphate-buffered saline and cell sample was assumed to be 1.33 and 1.36, respectively.

Note that the width of the wavelet mainly limits depth resolution. The wavelet duration in time is proportional to $P$; therefore, the value of $P^2$ affects the depth resolution when the contrast mechanism in the final image is the Brillouin frequency. When sectioning the two-layered sample consisting of phosphate-buffered saline ($f_B$ = 10 GHz) and $SiO_2$ ($f_B$ = 44 GHz) as in Figure 3, sampled phonon wavelength ($\lambda_{phonon} = \lambda/2n$) is 150 and 136 nm at $\lambda$ = 400 nm, respectively. As for the wavelet size, the temporal width of 10 and 44 GHz wavelets are 91 and 21 ps for $P^2 = 5$ (Figure S11a and S11b, Supporting Information). From the calculated acoustic velocity of phosphate-buffered saline ($v = 1.5$ km/s) and SiO$_2$ ($v = 5.9$ km/s), the spatial width of those wavelets is 137 and 124 nm. For $P^2 = 60$, the temporal width of 10 and 44 GHz wavelets are 288 and 66 ps (Figure S11c and S11d, Supporting Information), respectively. The spatial width of those wavelets is 433 and 389 nm. These values are comparable to or larger than the sampled phonon wavelength. Therefore, depth resolution is limited by the width of the wavelet. Moreover, depth resolution is traded for frequency resolution, increasing depth resolution decreases frequency resolution (Figure S11e and S11f, Supporting Information).

### 4.5. Thin film preparation

The chromium, titanium, and silicon nitride thin layers were deposited on a 100 μm thick glass substrate (0111650, Paul Marienfeld, DE) by a sputtering system (iMiller CFS4EP, Shibaura Mechatronics Corp., JP). The silicon dioxide layer was deposited by a different sputtering system (SIH-450, Ulvac Inc., JP). We prepared the SiO$_2$ step on the titanium film by partially masking the titanium layer with 500 μm thick silicon before depositing SiO$_2$.

### 4.6. Biological cell sample preparation

HeLa cells were cultured on an EtOH-sterilized 300 nm Ti film deposited on a 100 μm thick glass substrate for 24 h. The cells were fixed with 4% (w/v) paraformaldehyde solution (161-20141, Fujifilm Wako, JP) for 15 min at room temperature. After washing with distilled water, the cells were exposed to air.

## Acknowledgements

The authors thank Ayako Mizushima (the University of Tokyo) for assisting with the fabrication of the thin-film samples at the Ultrafine Lithography Nano-Measurement Center at the University of Tokyo, and Mitsuo Takeda (Utsunomiya University) for discussions on the experiments. We would like to thank Editage (www.editage.com) for English language editing. This work was supported in part by JST PRESTO (JPMJPR1902) and the Nanotechnology Platform of the Ministry of Education, Culture, Sports, Science and Technology (JPMXP09F20UT0110).

## Conflict of interests

The authors declare no competing financial interests.

## Keywords

Brillouin microscopy, phonon spectroscopy, chirped pulse spectroscopy, mechanical imaging

# Figures

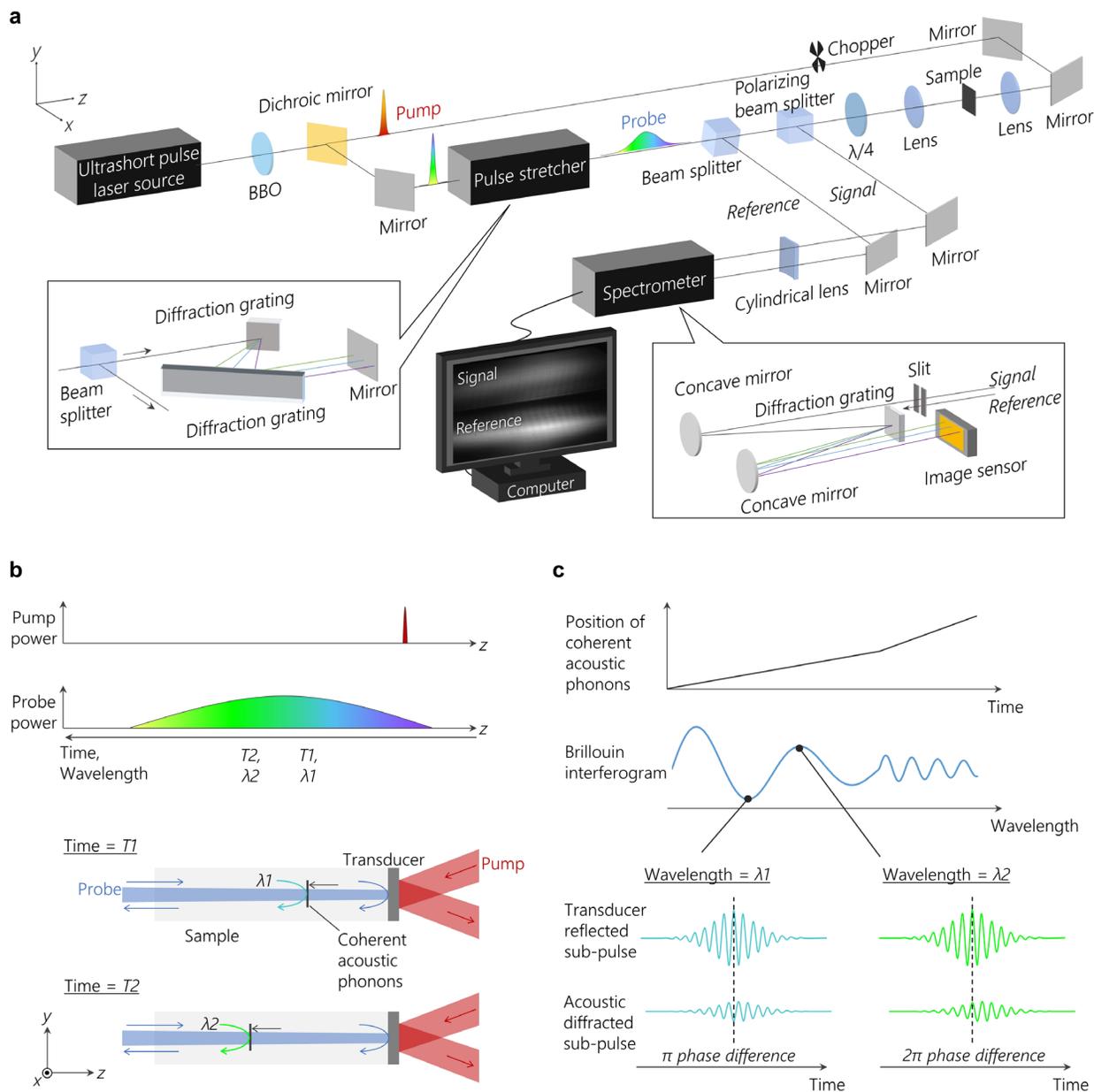

**Figure 1.** Schematic and conceptual illustration of dispersive coherent Brillouin scattering spectroscopy. a) Schematic of the experimental setup. A near-infrared femtosecond laser pulse is loosely focused on a transducer surface to excite coherent acoustic phonons. A frequency-doubled laser pulse is stretched in the grating pair (left inset) and directed to a spectrometer (right inset) as a signal after the sample is illuminated. A fraction of the chirped laser pulse that bypasses the sample is also directed into the spectrometer as a reference (the light path is not shown in the right inset) for balanced detection. b) Principle of dispersive coherent Brillouin scattering spectroscopy. Each sub-pulse of the chirped laser pulse (of wavelengths $\lambda_1$ and $\lambda_2$) is diffracted by the coherent acoustic phonons at different timings ($T_1$ and $T_2$) and in different positions. c) The spectroscopic heterodyning of chirped laser pulse yields the interferogram in a frequency domain that beats at Brillouin frequency. This beat is caused by the phase difference between the diffracted chirped laser pulse from coherent acoustic phonons and the reflected chirped laser pulse from the transducer surface.

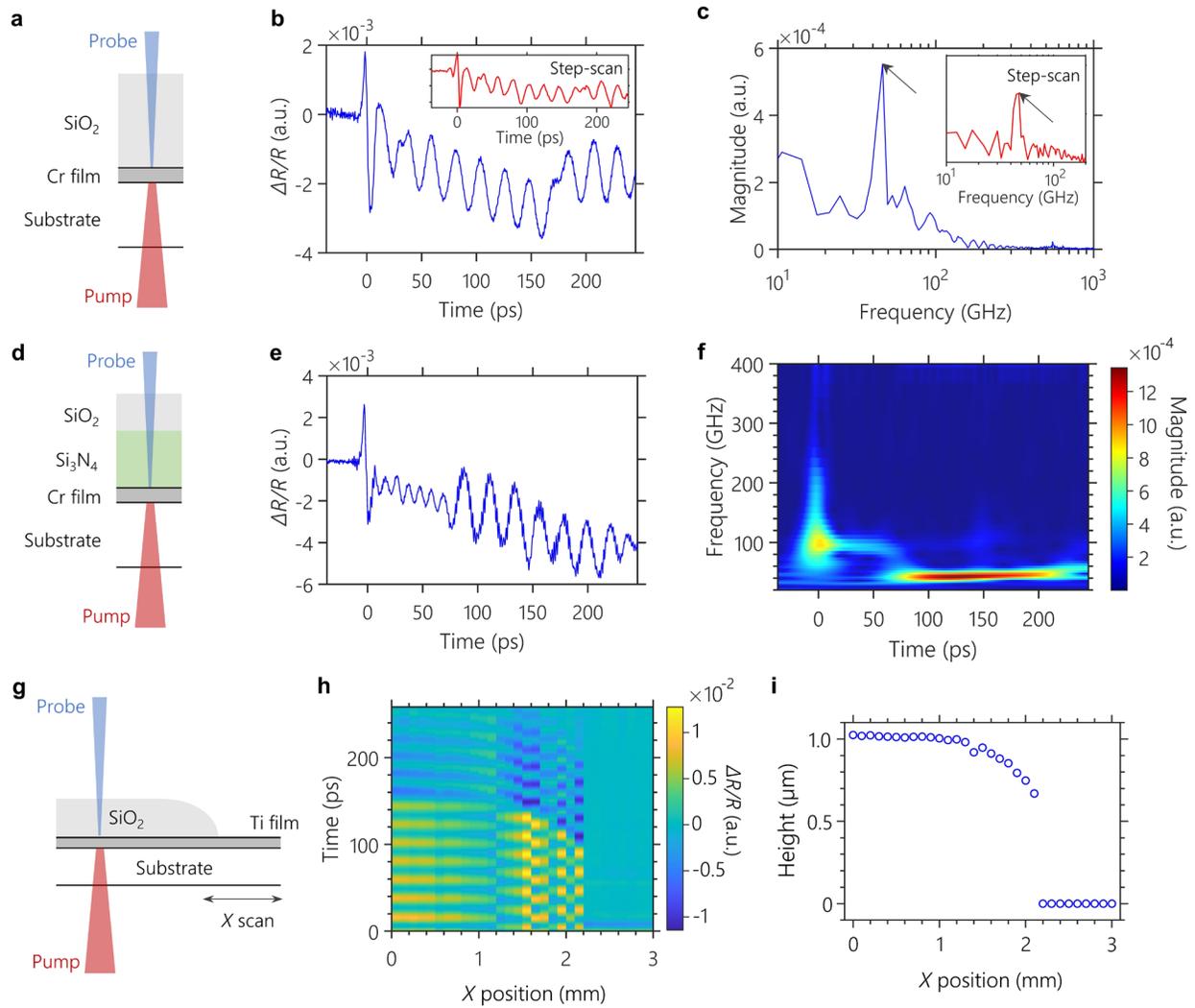

**Figure 2.** Basic performance of dispersive coherent Brillouin scattering spectroscopy. a) Schematic of the experiment with a 1 μm $SiO_2$ film sputtered on a 100 nm chromium film that launches coherent acoustic phonons through the transient absorption of an optical pump pulse (1.2 mJ/cm$^2$). The pump pulse excites the coherent acoustic phonons from the glass substrate side, which plays the role of the transducer. A chirped probe pulse then detects acoustic propagation in $SiO_2$. b) Brillouin oscillations acquired by dispersive heterodyne Brillouin spectroscopy (blue) and time-domain Brillouin spectroscopy with step-scan method (red). c) The Fourier transform of the measured Brillouin oscillations indicates a peak at 44 GHz. d) Schematic of the depth profiling experiment with the 700 nm $Si_3N_4$ film and 500 nm $SiO_2$ film sputtered on a 100 nm chromium

film. e) Recorded Brillouin oscillations and f) the corresponding time evolution of the Brillouin frequency indicating peaks around 90 GHz and 44 GHz. g) Schematic of the sample with 1 μm $SiO_2$ step created on a 300 nm titanium film. h) Recorded Brillouin oscillations and i) $SiO_2$ height at each scanning position.

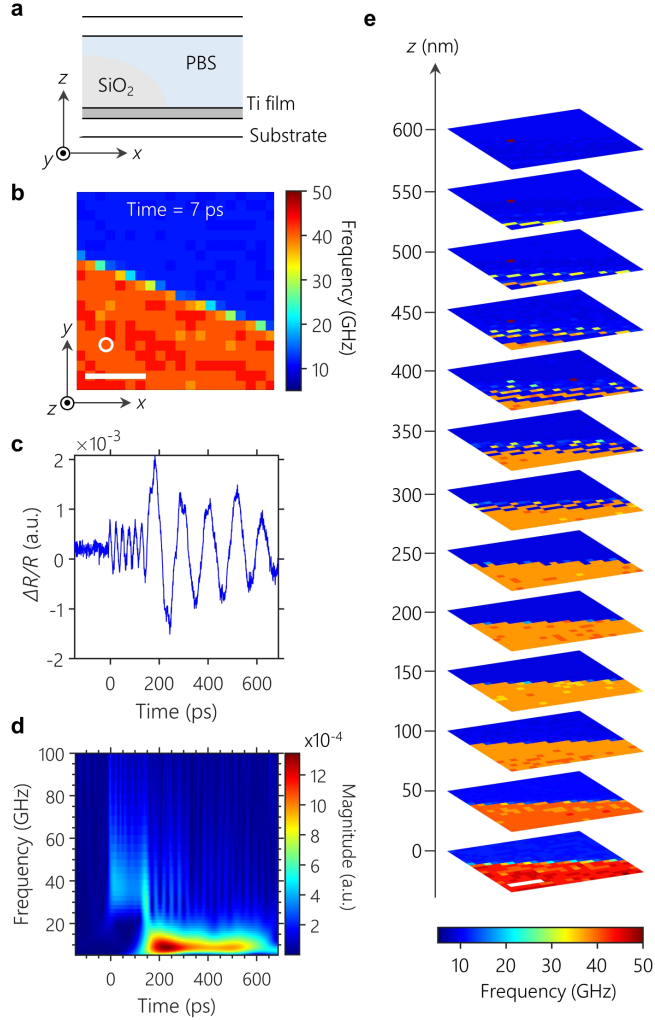

**Figure 3.** Three-dimensional dispersive coherent Brillouin scattering microscopy of thin film sample. a) The sample is phosphate-buffered saline squeezed between two flat glass substrates; one of them is a 300 nm titanium film with a 1 μm SiO2 step. b) Representative cross-sectional Brillouin frequency image of the sample at 7 ps, showing a signal from SiO$_2$ (red) and phosphate-buffered saline (blue). The difference in the Brillouin frequencies of the two materials produces a high contrast in the Brillouin frequency image. c) Recorded Brillouin oscillations and d) corresponding time evolution of the Brillouin frequency at the position identified as a white circle in the Brillouin frequency image. e) 3D Brillouin frequency images at different calculated depth position. Scale bar is 30 μm.

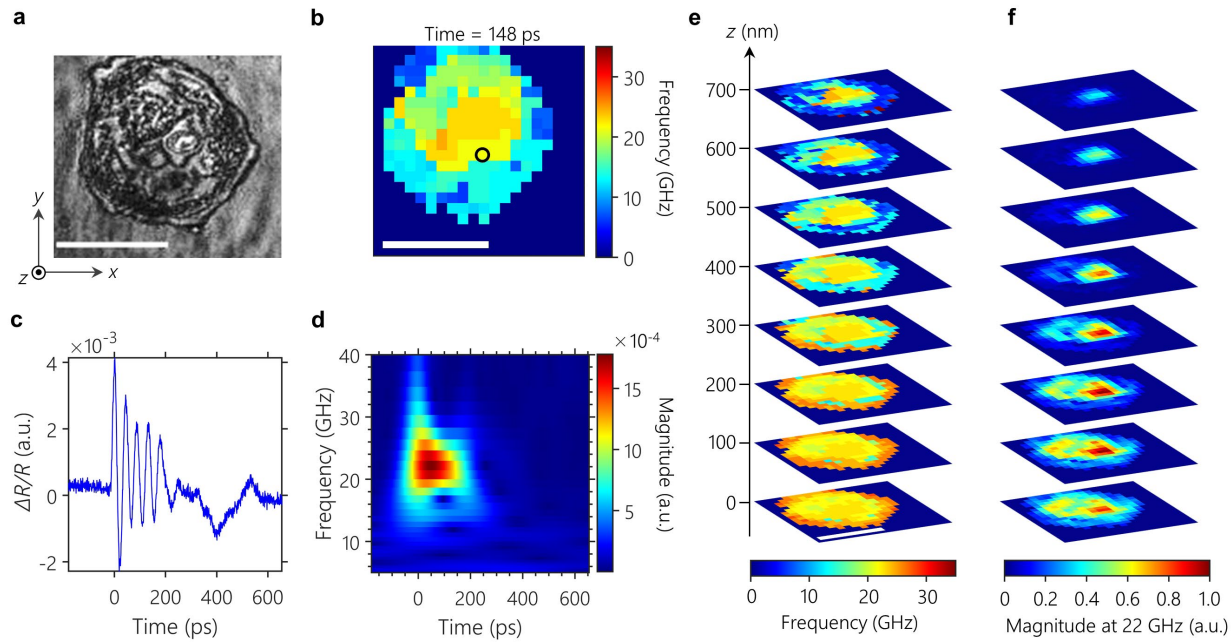

**Figure 4.** Three-dimensional dispersive coherent Brillouin scattering microscopy of biological cell. a) Brightfield image and b) Representative cross-sectional Brillouin frequency image of the HeLa cell at 148 ps. c) Recorded Brillouin oscillations and d) corresponding time evolution of the Brillouin frequency at the position identified as a black circle in Figure 4b. e) 3D Brillouin frequency images and f) 22 GHz spectral images at various calculated depth positions. The images are masked to remove the high frequency regions around the cells in the shallow area, which originate from the displacement of the Ti film surface. Scale bars are 30 μm.

Supporting Information

# Dispersive coherent Brillouin scattering spectroscopy

Ayumu Ishijima[1,2,*], Shinga Okabe[3], Ichiro Sakuma[1,3,4] & Keiichi Nakagawa[1,3]

*[1]Department of Precision Engineering, The University of Tokyo, Tokyo 113-8656, Japan.*
*[2]PRESTO, Japan Science and Technology Agency, Saitama 332-0012, Japan*
*[3]Department of Bioengineering, The University of Tokyo, Tokyo 113-8656, Japan.*
*[4]Medical Device Development and Regulation Research Center, The University of Tokyo, Tokyo 113-8656, Japan*
[*]ishijima@bmpe.t.u-tokyo.ac.jp

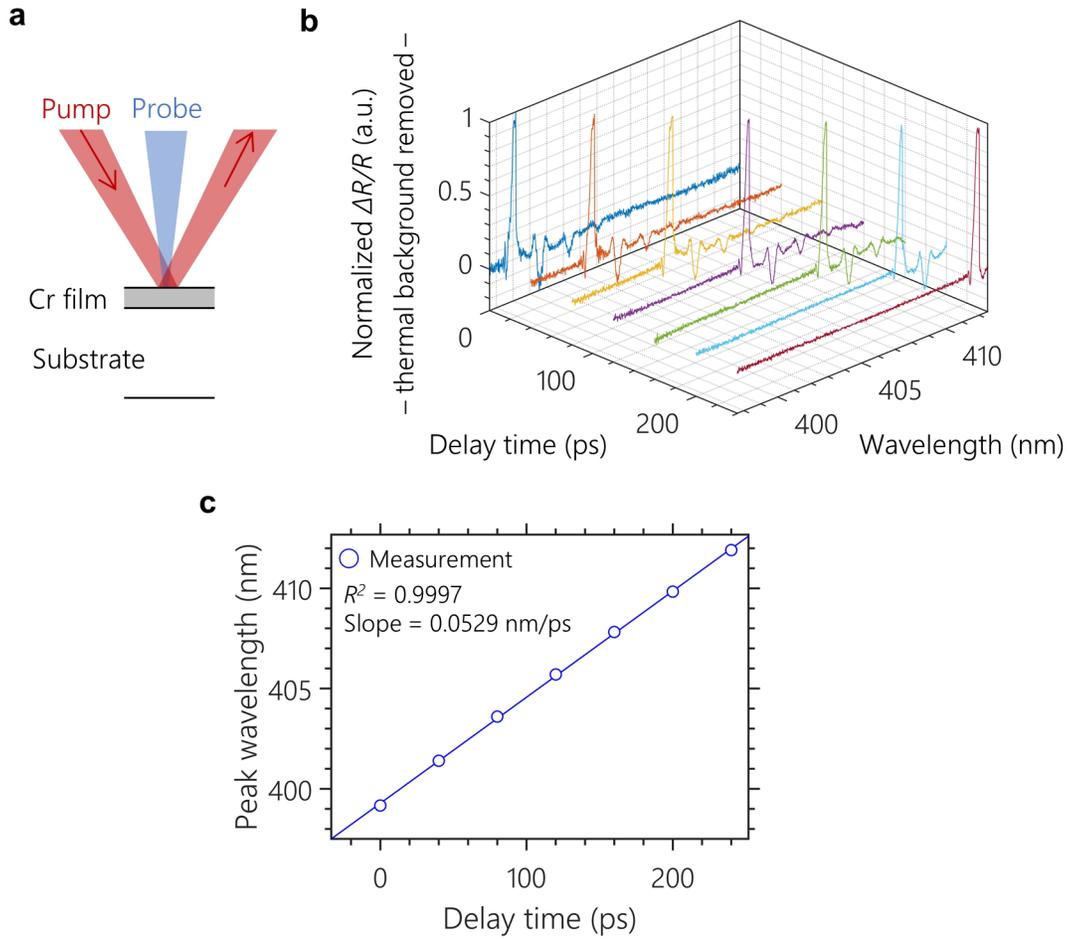

Figure S1 Characterization of the time window (282 ps). a) Schematic of the experiment with the 100 nm chromium film irradiating optical pump pulse. b) Transient reflectivity waveforms obtained by varying the pump delay time by stepping an optical delay line set in the pump arm. c) Plot of the wavelength at which the first transient reflectivity peak was detected against the delay time added to the pump light. Circles represent the measured data and the slope of the linear fitted line is 0.0529.

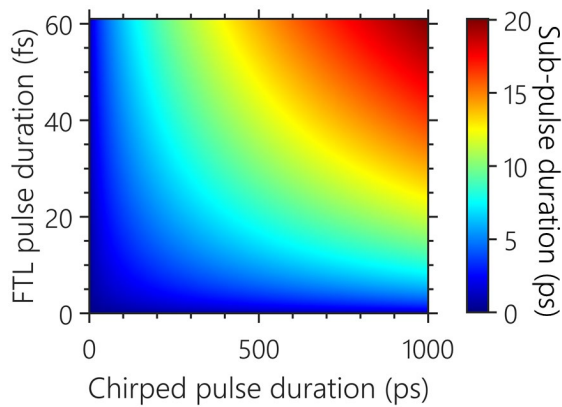

Figure S2 Dependence of the chirped pulse duration and initial Fourier transform-limited (FTL) pulse duration on the system's temporal resolution (the sub-pulse duration). The temporal resolution can be tuned to optimize the performance, e.g., by enlarging the bandwidth of the probe pulse.

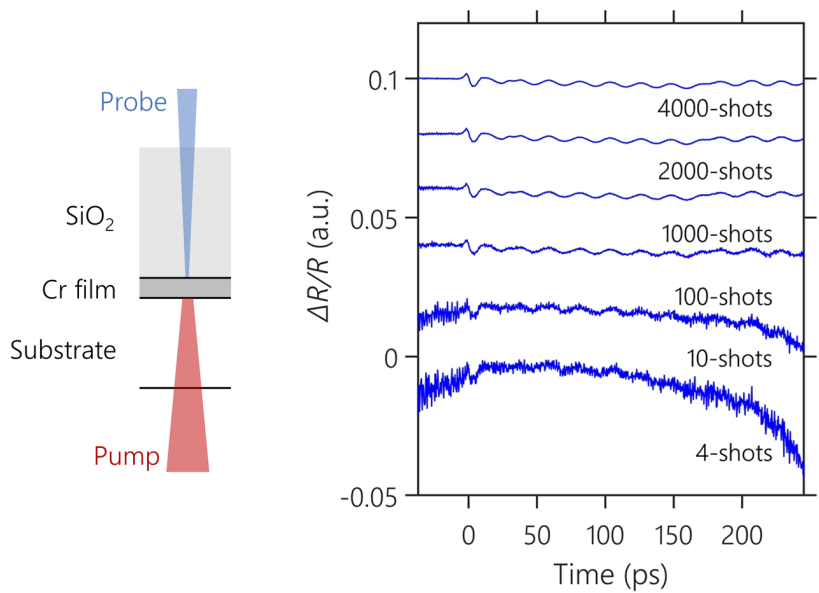

Figure S3 Recorded transient reflectivity waveforms of 1 μm $SiO_2$ film sputtered on a 100 nm chromium film with various averaging conditions. A series of transient reflectivity recorded from the same region of the sample displays the Brillouin oscillation of $SiO_2$ with an enhancement of signal-to-noise ratio level while increasing the number of shots to be averaged.

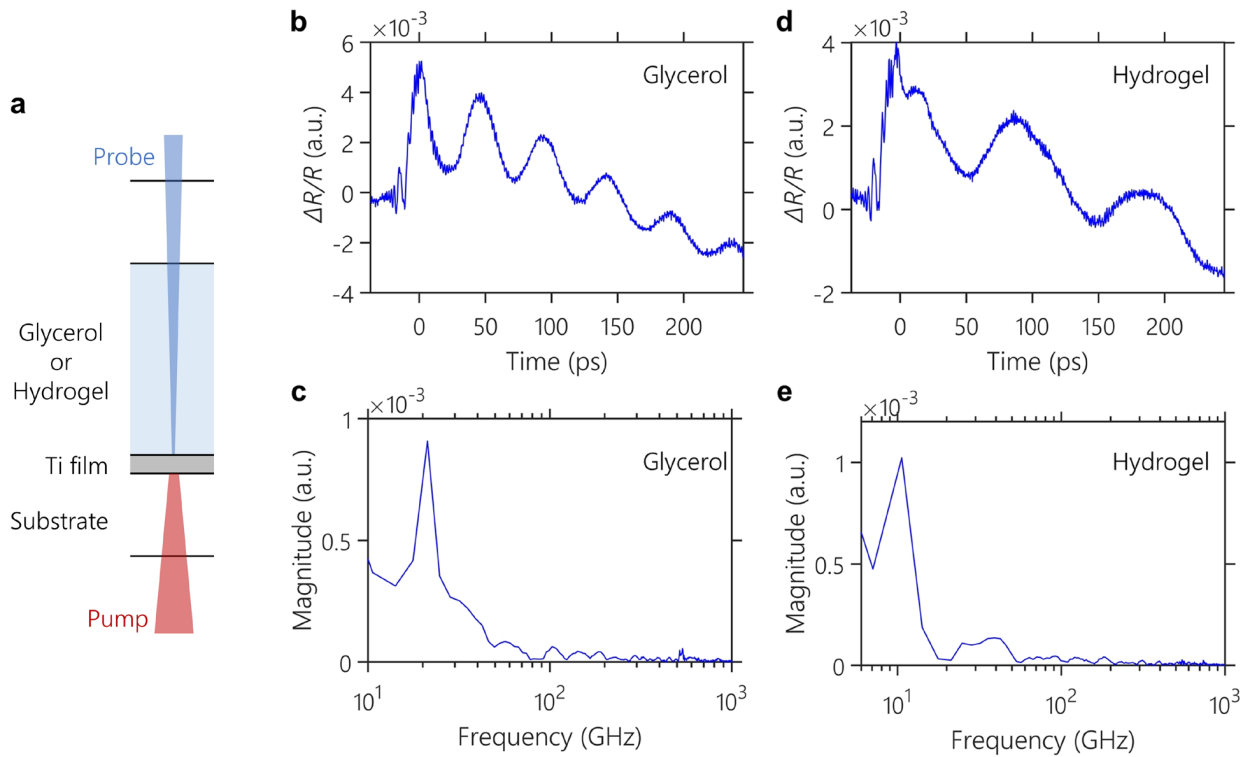

Figure S4 Dispersive coherent Brillouin scattering spectroscopy of liquid and gel. a) Schematic of the experiment with glycerol or hydrogel, which is squeezed between two flat glass substrates, one of them being coated with a titanium thin film. The hydrogel was made of 40% polyacrylamide solution, tetramethyl ethylenediamine, and distilled water at a volume ratio of 25:1:75. The hydrogel solution was fixed at 0.002% (v/v) of ammonium persulfate. b) and d) Recorded Brillouin oscillations and c) and e), corresponding spectra for glycerol and hydrogel displaying peaks at 20 and 10 GHz, respectively.

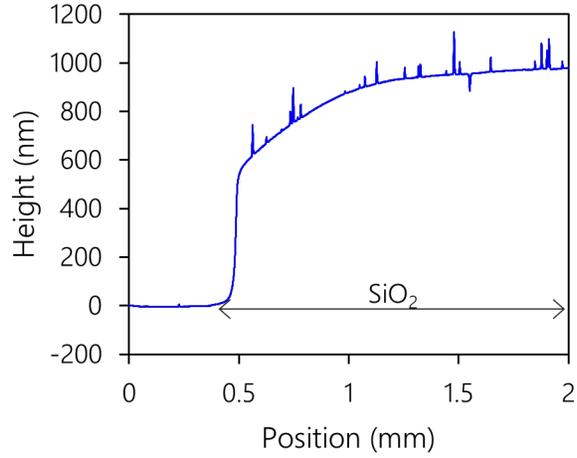

Figure S5 Height profile of SiO$_2$ step measured using a stylus profilometer. The peaks displayed in the profile are artifacts originating from the dust on the sample.

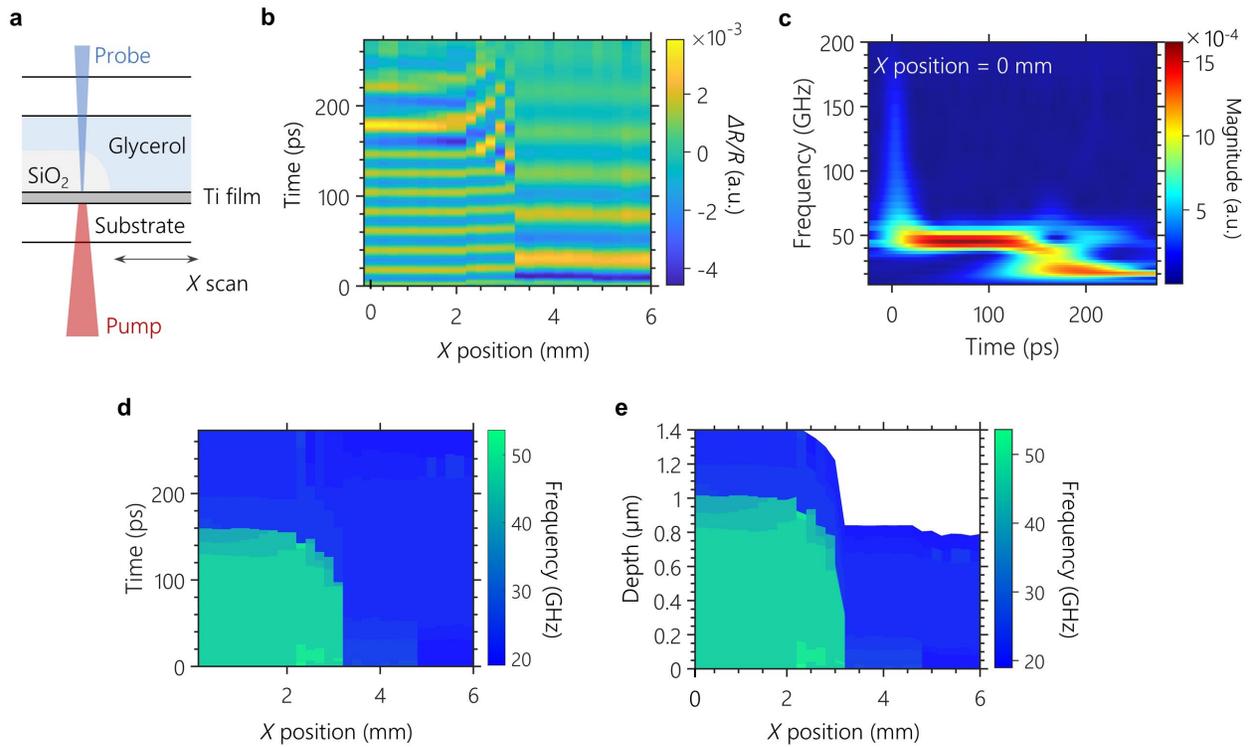

Figure S6 Cross-sectional Brillouin imaging by dispersive coherent Brillouin scattering spectroscopy. a) Schematic of the experiment with glycerol, which is squeezed between two flat glass substrates, one of them being coated with a 1 μm SiO$_2$ film and 300 nm titanium film. The sample was translated with a 200 µm-step over 6 mm to obtain the transient reflectivity map. A total of 31 waveforms were acquired, with 4000 shots averaging per step. b) Recorded Brillouin oscillations and c) corresponding time evolution of the Brillouin frequency at the lateral position of 0 mm. d) Brillouin frequency image of the sample showing a signal from $SiO_2$ (green) and glycerol (blue) with respect to the vertical axis of time and e) calculated depth position.

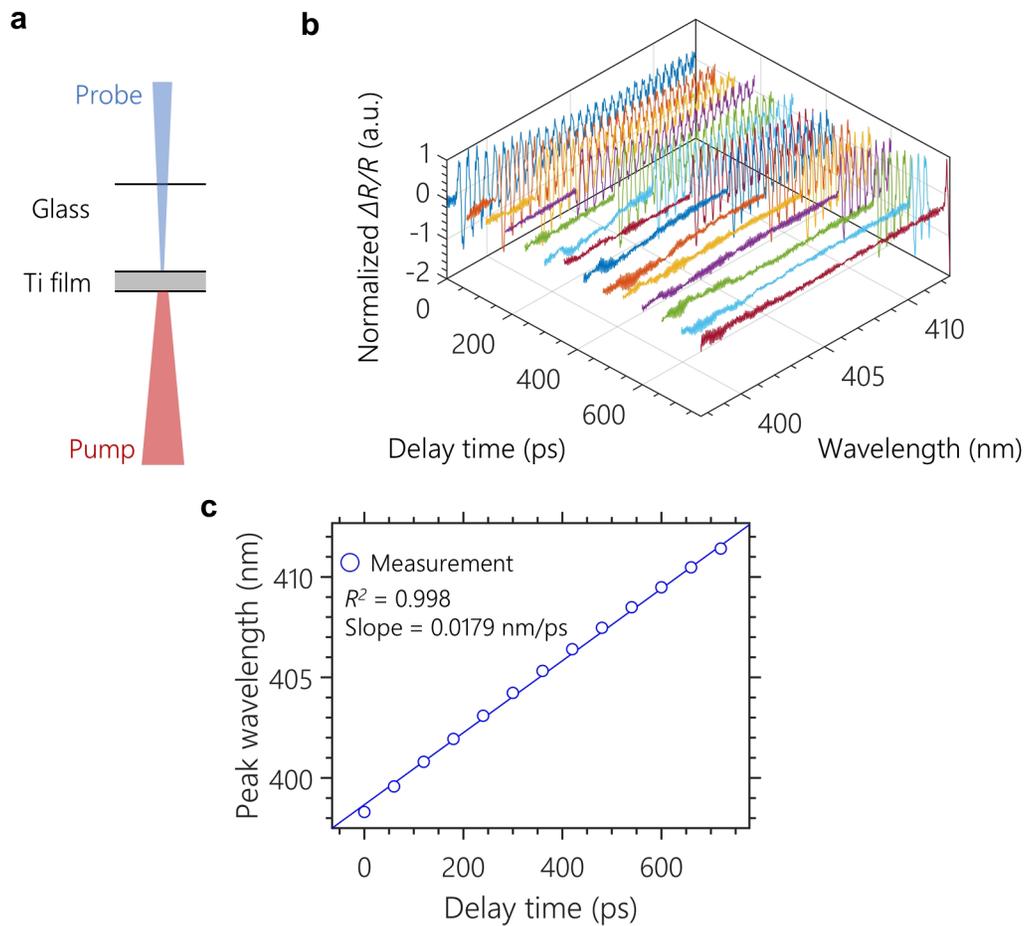

Figure S7 Characterization of the time window (844 ps). a) Schematic of the experiment with a 100 nm chromium film irradiating optical pump pulse. b) Transient reflectivity waveforms obtained by varying the pump delay time by stepping an optical delay line set in the pump arm. c) Plot of the wavelength at which the first transient reflectivity peak was detected against the delay

time added to the pump light. Circles represent measured data and the slope of linear fitted line is 0.0179.

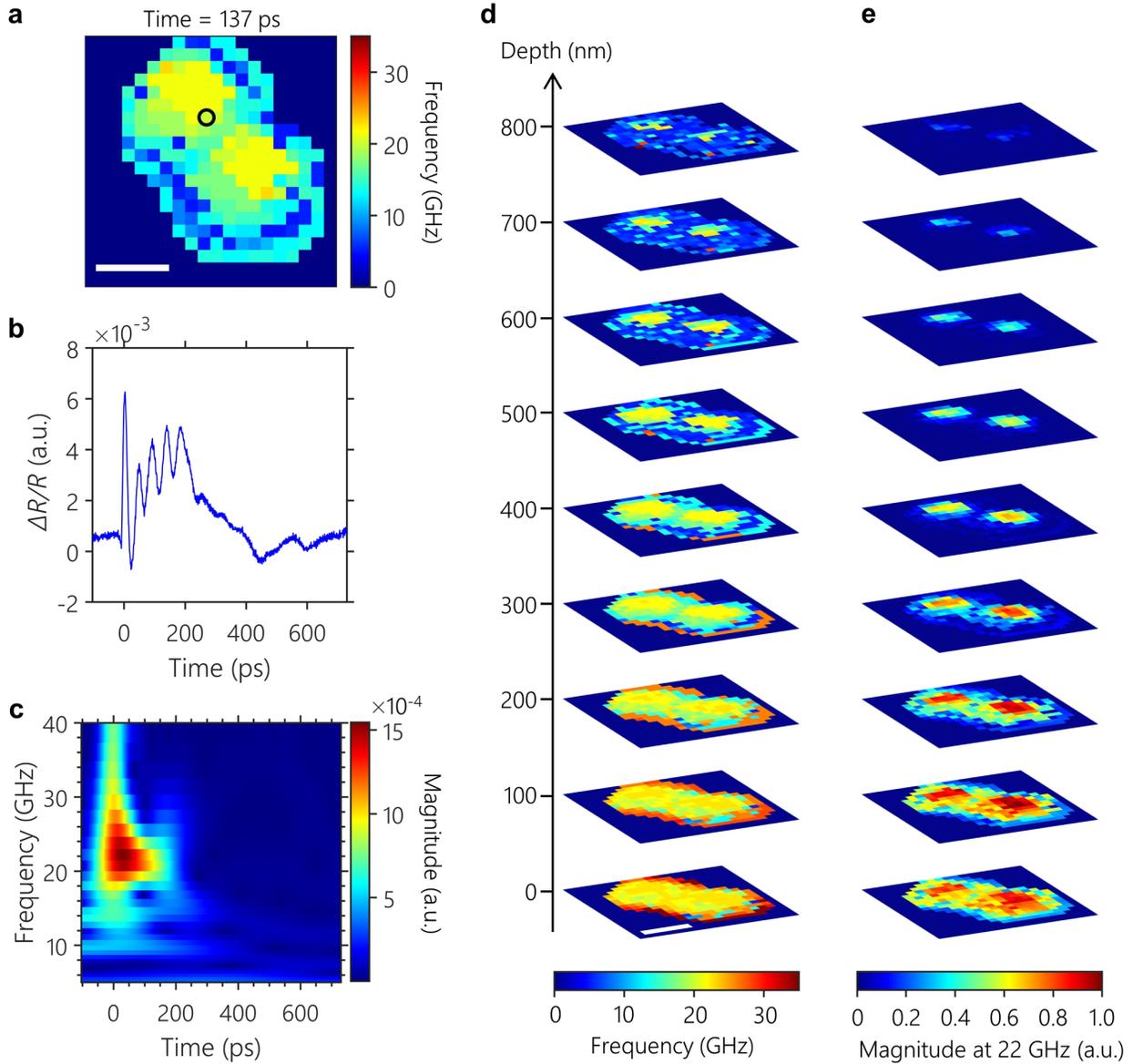

Figure S8 Three-dimensional Brillouin microscopy of biological cells different from Figure 4. a) Representative cross-sectional Brillouin frequency image of the HeLa cells at 137 ps. b) Recorded Brillouin oscillations and c) corresponding time evolution of the Brillouin frequency at the position identified as a black circle in the Brillouin frequency image. d) Three-dimensional Brillouin

frequency images and e) 22 GHz spectral images at various calculated depth positions. We can observe two cells adjacent to each other. Scale bars are 30 μm.

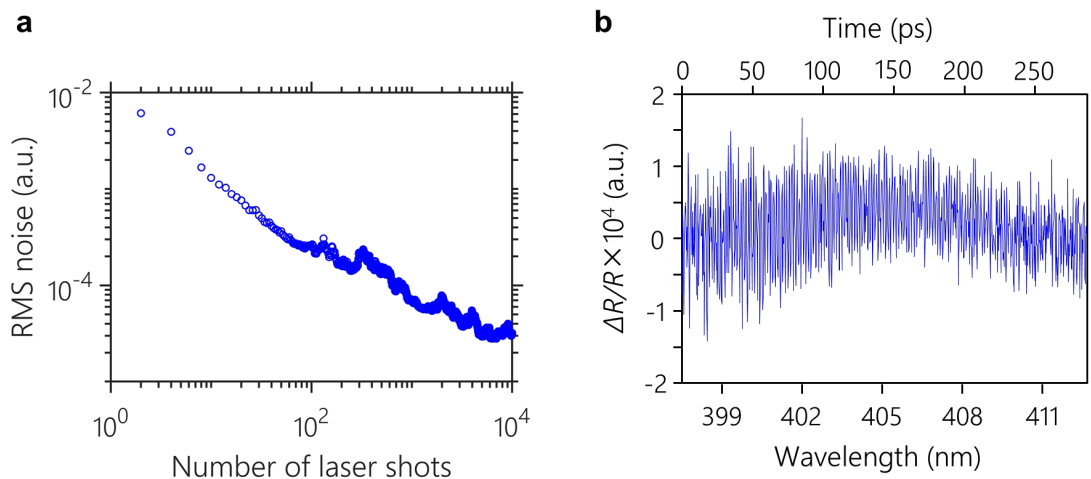

Figure S9 Characterization of the sensitivity. a) The root-mean-square (RMS) fluctuation value is defined by the standard deviation of the transient reflectivity waveform collected without irradiating pump pulses. As the number of averaged shots, $N$, increases, the RMS noise level decreases to approximately $N^{1/2}$ because the laser shot-to-shot fluctuations are random. b) Representative transient reflectivity waveform without pump pulses using 4000-shots of probe pulses for averaging.

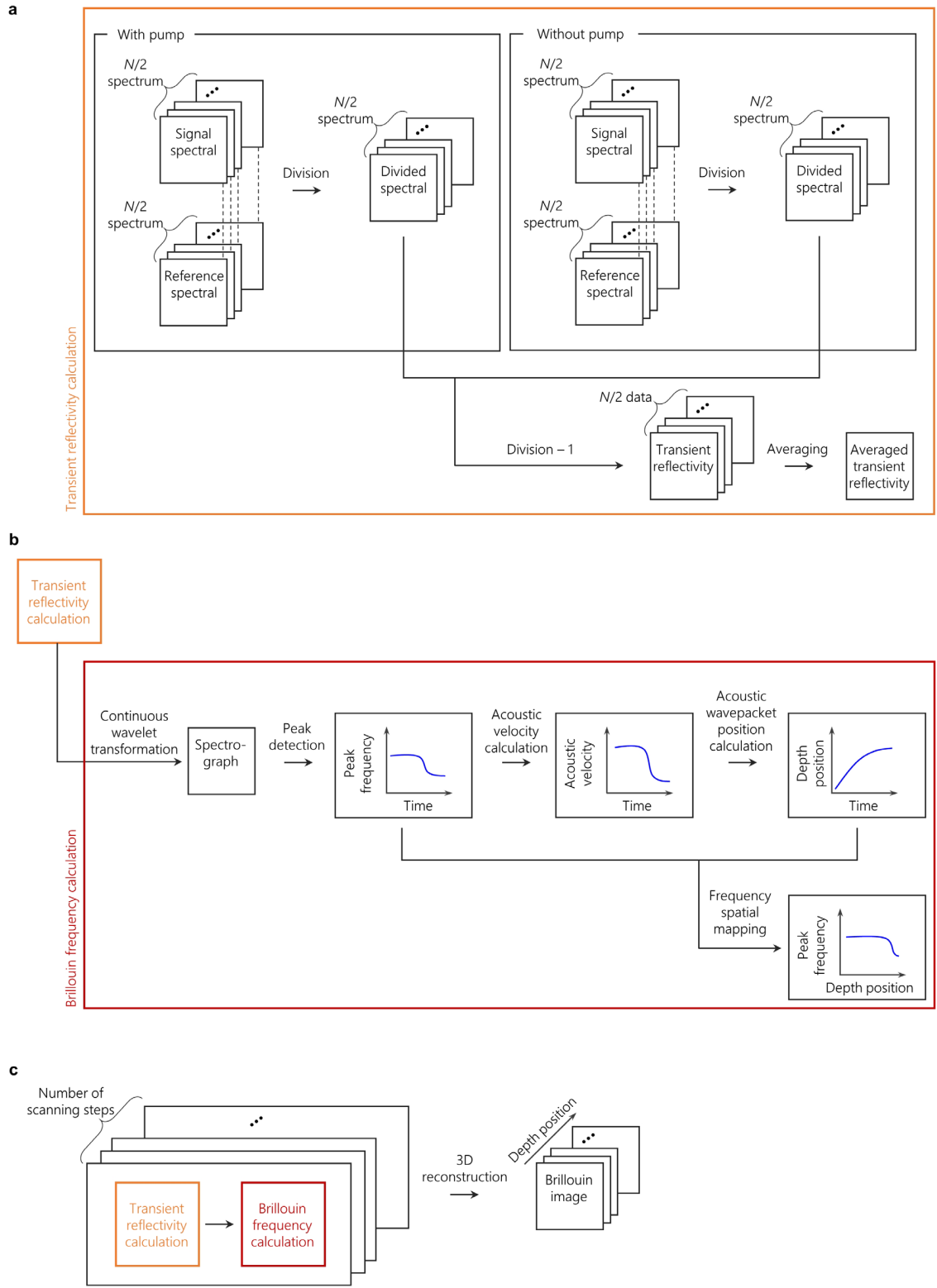

Figure S10 Calculation flow for a) transient reflectivity waveform, b) Brillouin frequency spatial mapping, and c) 3D Brillouin image reconstruction. *N* denotes the number of laser shots.

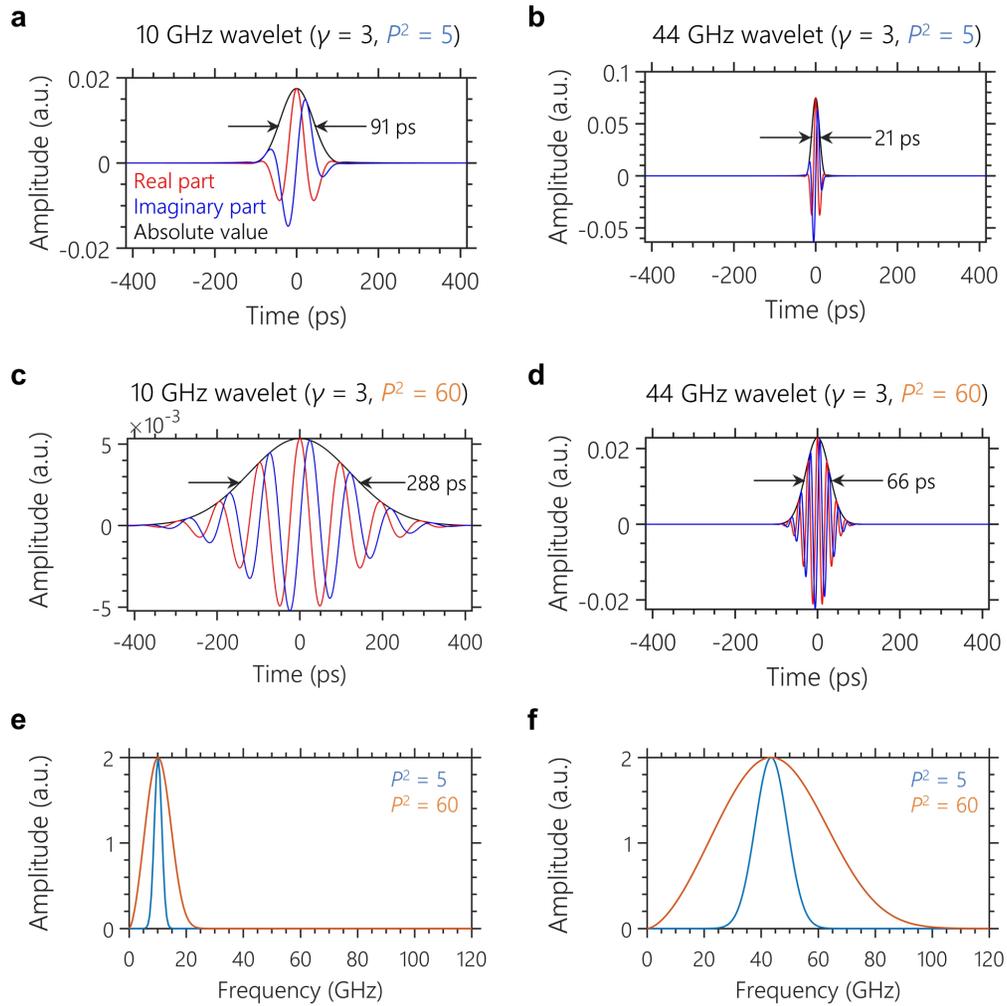

Figure S11 a) The time-domain wavelets for 10 and b) 44 GHz center frequencies for the time-bandwidth product $P^2$ of 5. c) The time-domain wavelets for 10 and d) 44 GHz center frequencies for the time-bandwidth product $P^2$ of 60. e) Frequency response of 10 and f) 44 GHz wavelets.